\documentclass[aps,prl,onecolumn,amsmath,amssymb,superscriptaddress]{revtex4}
\usepackage{graphicx}
\usepackage{amssymb}
\usepackage{natbib}
\usepackage{color} 
\usepackage{textcomp}
\usepackage[normalem]{ulem}
\usepackage[dvipsnames]{xcolor}

\newcommand{\hl}[1]{#1}

\newcommand{\hy}[1]{#1}

\newcommand{\bg}[1]{#1}

\newcommand{\CZO}{Ce$_2$Zr$_2$O$_7$}
\newcommand{\CSO}{Ce$_2$Sn$_2$O$_7$}

\newcommand{\half}{\ensuremath{\frac12}}
\newcommand{\bfq}{\mathbf{k}}

\begin{document}

\title{Magnetic field effects in an  octupolar quantum spin liquid candidate}

\author{Bin Gao}
\thanks{These authors made equal contributions to this work.}
\affiliation{Department of Physics and Astronomy, Rice University, Houston, Texas 77005, USA }

\author{Tong Chen}
\thanks{These authors made equal contributions to this work.}
\affiliation{Department of Physics and Astronomy, Rice University, Houston, Texas 77005, USA }

\author{Han Yan}
\affiliation{Department of Physics and Astronomy, Rice University, Houston, Texas 77005, USA }

\author{Chunruo Duan}
\affiliation{Department of Physics and Astronomy, Rice University, Houston, Texas 77005, USA }

\author{Chien-Lung Huang}
\affiliation{Department of Physics and Astronomy, Rice University, Houston, Texas 77005, USA }

\author{Xu Ping Yao}
\affiliation{Department of Physics and HKU-UCAS Joint Institute 
for Theoretical and Computational Physics at Hong Kong, The University of Hong Kong, Hong Kong, China}

\author{Feng Ye}
\affiliation{Neutron Scattering Division, Oak Ridge National Laboratory, Oak Ridge, Tennessee 37831, USA}

\author{Christian Balz}
\affiliation{ISIS Facility, STFC Rutherford-Appleton Laboratory, Didcot, OX11 0QX, UK}

\author{J. Ross Stewart}
\affiliation{ISIS Facility, STFC Rutherford-Appleton Laboratory, Didcot, OX11 0QX, UK}

\author{Kenji Nakajima}
\affiliation{
Neutron Science Section, Materials and Life Science Division, J-PARC Center, Tokai, Ibaraki 319-1195, Japan}

\author{Seiko Ohira-Kawamura}
\affiliation{
Neutron Science Section, Materials and Life Science Division, J-PARC Center, Tokai, Ibaraki 319-1195, Japan}

\author{Guangyong Xu}
\affiliation{NIST Center for Neutron Research, National Institute of Standards and Technology, Gaithersburg, Maryland 20899, USA}

\author{Xianghan Xu}
\affiliation{Rutgers Center for Emergent Materials and Department of Physics and Astronomy, Rutgers University, Piscataway, New Jersey 08854, USA}

\author{Sang-Wook Cheong}
\affiliation{Rutgers Center for Emergent Materials and Department of Physics and Astronomy, Rutgers University, Piscataway, New Jersey 08854, USA}

\author{Emilia Morosan}
\affiliation{Department of Physics and Astronomy, Rice University, Houston, Texas 77005, USA }

\author{Andriy H. Nevidomskyy}
\thanks{nevidomskyy@rice.edu}
\affiliation{Department of Physics and Astronomy, Rice University, Houston, Texas 77005, USA }

\author{Gang Chen}
\thanks{gangchen@hku.hk}
\affiliation{Department of Physics and HKU-UCAS Joint Institute 
for Theoretical and Computational Physics at Hong Kong, The University of Hong Kong, Hong Kong, China}

\author{Pengcheng Dai}
\thanks{pdai@rice.edu}
\affiliation{Department of Physics and Astronomy, Rice University, Houston, Texas 77005, USA }

\begin{abstract}
Quantum spin liquid (QSL) is a disordered state of quantum-mechanically entangled spins commonly arising from frustrated magnetic dipolar interactions. 
However, QSL in some pyrochlore magnets can also come from frustrated magnetic octupolar interactions.
Although the key signature for both dipolar and octupolar interaction-driven QSL is the presence of a spin excitation continuum (spinons) arising from the spin quantum number fractionalization, an external magnetic field-induced ferromagnetic order will transform the spinons into conventional spin waves in a dipolar QSL. 
By contrast, in an octupole QSL, the spin waves carry octupole moments that do not couple, in the leading order, to external magnetic field or to neutron moments but will contribute to the field dependence of the heat capacity.
Here we use neutron scattering to show that the application of a large external magnetic field to Ce$_2$Zr$_2$O$_7$, an octupolar QSL candidate, induces an Anderson-Higgs transition by condensing the spinons into a static ferromagnetic ordered state 
with octupolar spin waves invisible to neutrons but contributing to the heat capacity.  Our theoretical calculations also provide a microscopic, \hl{qualitative} understanding for the presence of 
 octupole scattering at large wavevectors in Ce$_2$Sn$_2$O$_7$ pyrochlore, and its absence in Ce$_2$Zr$_2$O$_7$. Therefore, our results identify Ce$_2$Zr$_2$O$_7$ as \hl{a strong candidate for} an octupolar $U(1)$ QSL, establishing that frustrated magnetic octupolar interactions are responsible for QSL properties in Ce-based pyrochlore magnets.
\end{abstract}

\maketitle

\section{introduction}
A quantum spin liquid (QSL) is a disordered state of entangled quantum spins that does not exhibit any long-range magnetic order in the zero-temperature limit \cite{anderson1973,balents2010,zhou2017,Savary2017,broholm2020}. 
Originally proposed by Anderson as the ground state for a system of ${S=1/2}$ spins on the two-dimensional (2D) triangular lattice with antiferromagnetic nearest neighbor interactions~\cite{anderson1973}, many other geometrically frustrated lattices have been suggested to harbor QSL~\cite{balents2010}. 
 A key experimental signature of a QSL is the presence of deconfined spinons, fractionalized quasiparticles carrying spin-\half, that can be observed by inelastic neutron scattering as a broad spin excitation continuum around the Brillouin zone boundary in the reciprocal space \cite{broholm2020}.
In typical candidate systems such as kagom\'{e} \cite{Han2012}, triangular \cite{jun2016,mourigal2017,pldai2021}, distorted kagom\'{e} bilayers \cite{Balz2016}, and pyrochlore \cite{Hallas2018} lattices, the observed spin excitation continuum may arise from the entanglement of magnetic dipolar interactions of (effective) 
${S=1/2}$ quantum spins 
 on geometrically frustrated lattices. 
For example, in rare-earth pyrochlores, the magnetic dipole-dipole (or anti-ferromagnetic) interactions of Ising-like moments decorated on a lattice of corner-sharing tetrahedra [Figs. 1(a-d)] lead to the well-known ‘2-in/2-out’ spin ice arrangement \cite{Bramwell2001,Castelnovo2012,Gingras2014}.
As a consequence, a QSL state can emerge in the so-called pyrochlore quantum spin ice regime characterized by the $U(1)$ quantum electrodynamics with emergent photon-like gapless excitations and gapped magnetic and electric charges 
 \cite{Hermele2004,Savary2012,Gingras2014,Benton2012,Chen2016,Chen2017s,Chen2017}.

\begin{figure}[t]
\includegraphics[width=0.8\columnwidth]{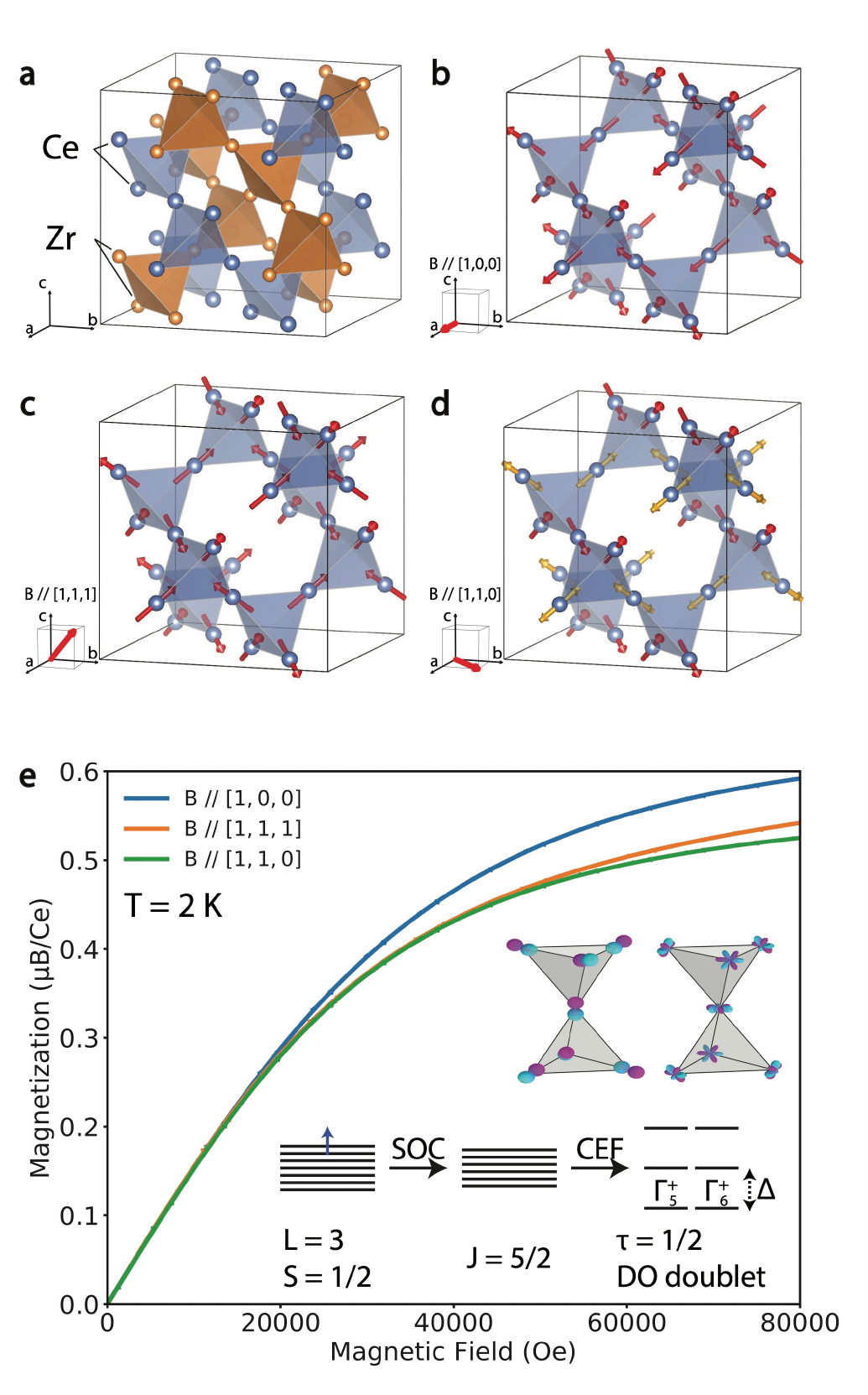}
\caption{
(a) A schematic of the structure of Ce$_2$Zr$_2$O$_7$ with lattice parameters a = b = c = 10.71 \AA. The blue ions are magnetic Ce$^{3+}$ (A site) and the brown ions are non-magnetic Zr$^{4+}$ (B site). 
(b)-(d) Schematics of field-induced magnetic structures of Ce2$_2$Zr$_2$O$_7$ in magnetic fields along the $[1,0,0]$, $[1,1,1]$, and $[1,1,0]$ direction, respectively. 
(e) Magnetization ($M$) as a function of applied magnetic field ($H$) at 2 K for three different field directions. Inset: Schematics of dipolar and octupolar ice phases, and the dipole-octupole doublet CEF ground state in Ce2$_2$Zr$_2$O$_7$. 
The CEF gap was reported to be 55 meV.
}
\label{Fig1}
\end{figure}

\begin{figure}[t]
\includegraphics[width=0.8\columnwidth]{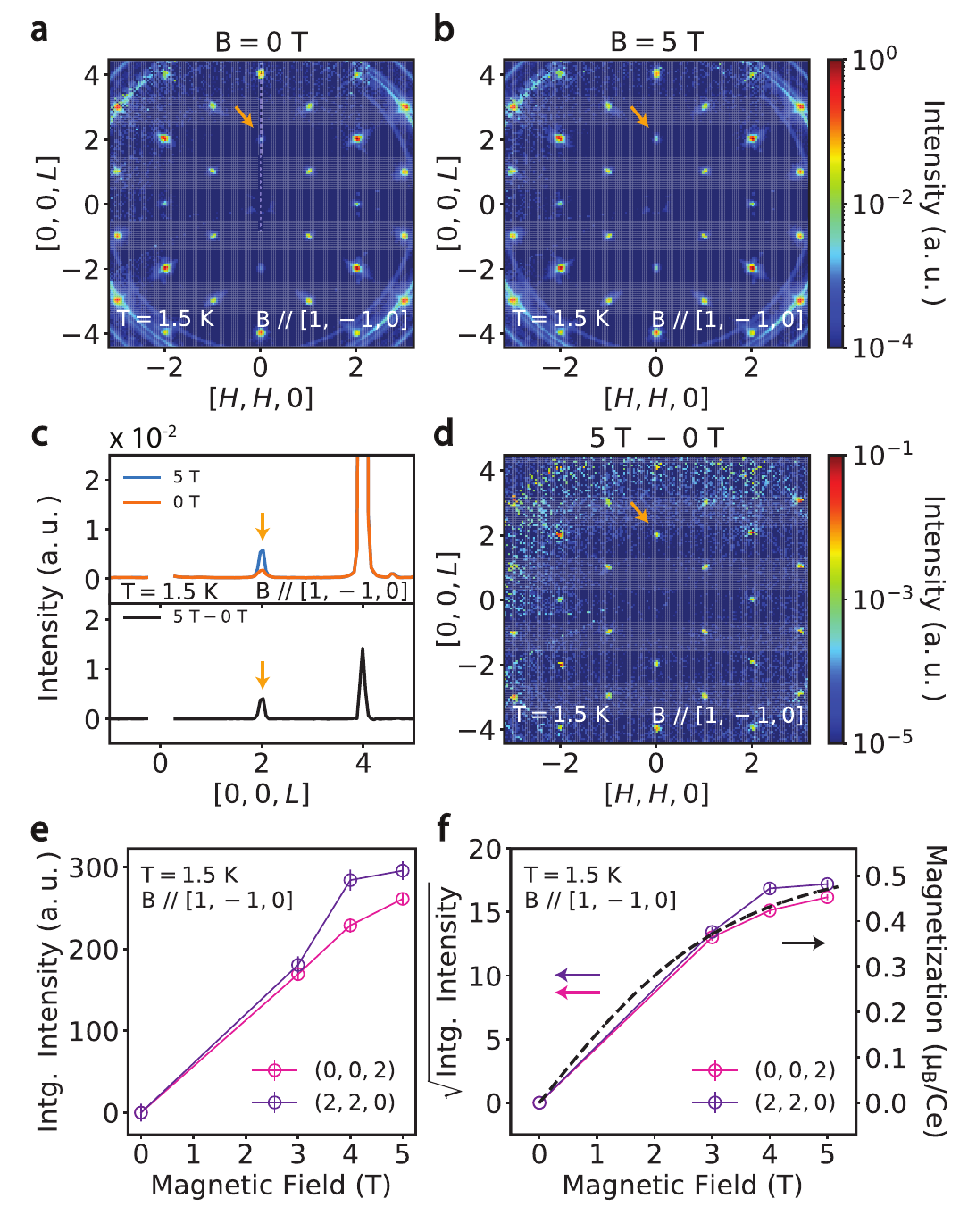}
\caption{
(a),(b) Wavevector dependence of the diffuse neutron scattering at 1.5 K in zero and 5 T magnetic field along the $[1,-1,0]$ direction. 
The dashed line in (a) indicates the wavevector direction for cuts in (c), and the orange arrows point to the (0,0,2) peak. 
In zero field, scattering at (0,0,2) is from multiple scattering. 
(c) Wavevector cuts along the $[0,0,L]$ direction at 1.5 K in zero and 5 T magnetic fields (upper panel), and their difference (lower panel). 
(d) Difference between the wavevector dependence of the diffuse scattering at 1.5 K in zero and 5 T magnetic field. 
(e) Integrated magnetic intensity of (0,0,2) and (2,2,0) Bragg peaks as a function of an applied magnetic field at 1.5 K. The intensity of the Bragg peaks in zero field was subtracted as background. 
(f) The square root of the integrated magnetic intensity as a function of a magnetic field in (e) overlapped with Magnetization data shown in Fig. 1(e). \bg{Data are from CORELLI spectrometer.}
}
\label{Fig2}
\end{figure}

On the other hand, $f$-electron ions in pyrochlore lattice~\cite{GingrasRMP} can carry multipole degrees of freedom of higher rank than dipoles, resulting in more complicated magnetic interactions \cite{Santini2009,GingrasTbTiO}. 
However, very few pyrochlores, for example Tb$_2$Ti$_2$O$_7$ \cite{GingrasTbTiO,Guitteny,Takatsu,Fritsch2013,Rule2006,fennell2012,HCao2008}, have been identified as high-order multipole systems. 
\hl{In particular, the strong magnetoelectric coupling in Tb$_2$Ti$_2$O$_7$ can lift the degeneracy of a QSL state and a magnetic field along the $[1,1,0]$ direction 
can induce the 2-in/2-out spin order \cite{Khomskii2012,Jaubert2015}.}
Recently, Ce$_2$Sn$_2$O$_7$ \cite{Sibille2015,Sibille2020} and its isoelectronic sister compound Ce$_2$Zr$_2$O$_7$ \cite{Gaudet2019,Gao2019} [denoted as Ce$_2T_2$O$_7$ ($T=$Sn, Zr)] have been proposed as a unique three-dimensional (3D) pyrochlore lattice QSL material with minimum magnetic and non-magnetic chemical disorder. 
For the Ce-based pyrochlore structure in the $Fd\bar{3}m$ space group, the crystal electric field (CEF) potential from the eight oxygen anions (the $D_{3d}$ crystal field) will split the Ce$^{3+}$ ion with an odd number of $4f$ electrons 
($4f^1$, $^2F_{5/2}$) into three Kramers doublets
 [inset of Fig. 1(e)] \cite{Gingras2014}.  
Since each state in the Kramers doublet is a 1D irreducible representation $\Gamma_5^{+}$ and $\Gamma_6^{+}$ of the $D_{3d}$ double point group, the pseudo-spins of the ground state doublet can transform like degenerate magnetic dipoles and magnetic octupoles (rank-3 multipoles), thus dubbed dipole-octupole doublet [inset of Fig. 1(e)] \cite{Huang2014,Li2017,Yao2020}. 
This is very different from the Kramers doublet of Yb$^{3+}$ ground state in Yb$_2$Ti$_2$O$_7$, where the doublet forms a 2D irreducible representation $\Gamma_4^{+}$ of the $D_{3d}$ (double) point group with an effective spin-\half\ local moments \cite{Gingras2014,Ross2011}. 
In Yb$_2$Ti$_2$O$_7$, magnetic interactions are dominated by classical ${S=1/2}$ dipole-dipole interaction acting as ferromagnetic first-neighbor couplings, resulting in a QSL state in the spin ice regime \cite{Gingras2014,Savary2012}. 
The application of a large magnetic field can drive such a quantum spin ice
 into a field-polarized ferromagnet via spinon condensation, where the elementary excitations are conventional transverse spin waves and can be probed by inelastic neutron scattering measurements to extract the microscopic exchange parameters in the spin Hamiltonian \cite{Ross2011,Coldea2017}.

In the case of the Ce-based pyrochlores with the dipole-octupole doublet, the ground state can support two symmetry-enriched $U(1)$
QSLs: a dipolar QSL and an octupolar QSL, distinguished by the roles of the dipole and octupole components in each phase \cite{Huang2014,Li2017,Yao2020,Patri2020,Benton2020}. 
If the ground state of Ce-based pyrochlores is a dipolar QSL \cite{Gaudet2019}, it should behave like a quantum spin ice, and the application of a magnetic field should drive the system into a field-polarized ferromagnet where spin waves can be measured to determine the magnetic exchange couplings of the spin Hamiltonian, \hl{analogous to that of Yb$_2$Ti$_2$O$_7$ \cite{Ross2011,Coldea2017}}.  
However, if the ground state is an octupolar QSL \cite{Sibille2020}, the application of a large magnetic field should induce a field-driven Anderson-Higgs transition by condensing the spinons (spin excitation continuum) into a ferromagnet with
magnetic dipole moment $S^z$ polarized along the local $[1, 1, 1]$ direction of the tetrahedron \cite{Huang2014,Li2017}.
Since the magnetic octupolar moments $S^y$ \hl{
	 couple very weakly}  to the 
external magnetic field and thus cannot be seen by neutrons \hl{in practice}, at least to 
leading order \hl{of intensity in the small wavevector regime}, 
neutron scattering can only detect 
$\langle S^z S^z \rangle$
spin correlations. 
As a result, the observed spinon continuum in zero field should 
decrease with increasing field and there should be no \hl{observable} transverse spin waves 
($\langle S^x S^x \rangle$ and $\langle S^y S^y \rangle$ spin correlations) 
\hl{by inelastic neutron scattering}
in a field-driven, fully polarized ferromagnetic state \cite{Huang2014,Li2017}.
\hl{However, such magnon excitations physically exist, and are detectable indirectly in specific heat measurements.
The contrast of specific heat and neutron measurements can thus be a strong evidence of the octupolar nature of the spin system.}

Furthermore, it was argued that the octupoles ($S^y$ components) should be detectable via the pronounced enhancement of 
neutron diffusive scattering at low temperatures at large wavevectors.
Such enhancement has been observed on powder samples of Ce$_2$Sn$_2$O$_7$  \cite{Sibille2020}, suggesting that the dominant interaction must occur between the octupolar $S^y$ components. This in turn supports the notion of the octupolar U(1) QSL -- a coherent quantum state formed out of the  manifold of $S^y-$ice states.  However, similar measurements on single crystals will provide \hl{further} important information on the angular dependence of the octupolar diffuse scattering \hl{not available in a powder measurements} \cite{Sibille2020}.
It is thus of great theoretical and experimental interest to examine the diffuse neutron scattering in single crystals of Ce$_2T_2$O$_7$ family.

\begin{figure}[t]
\includegraphics[width=0.8\columnwidth]{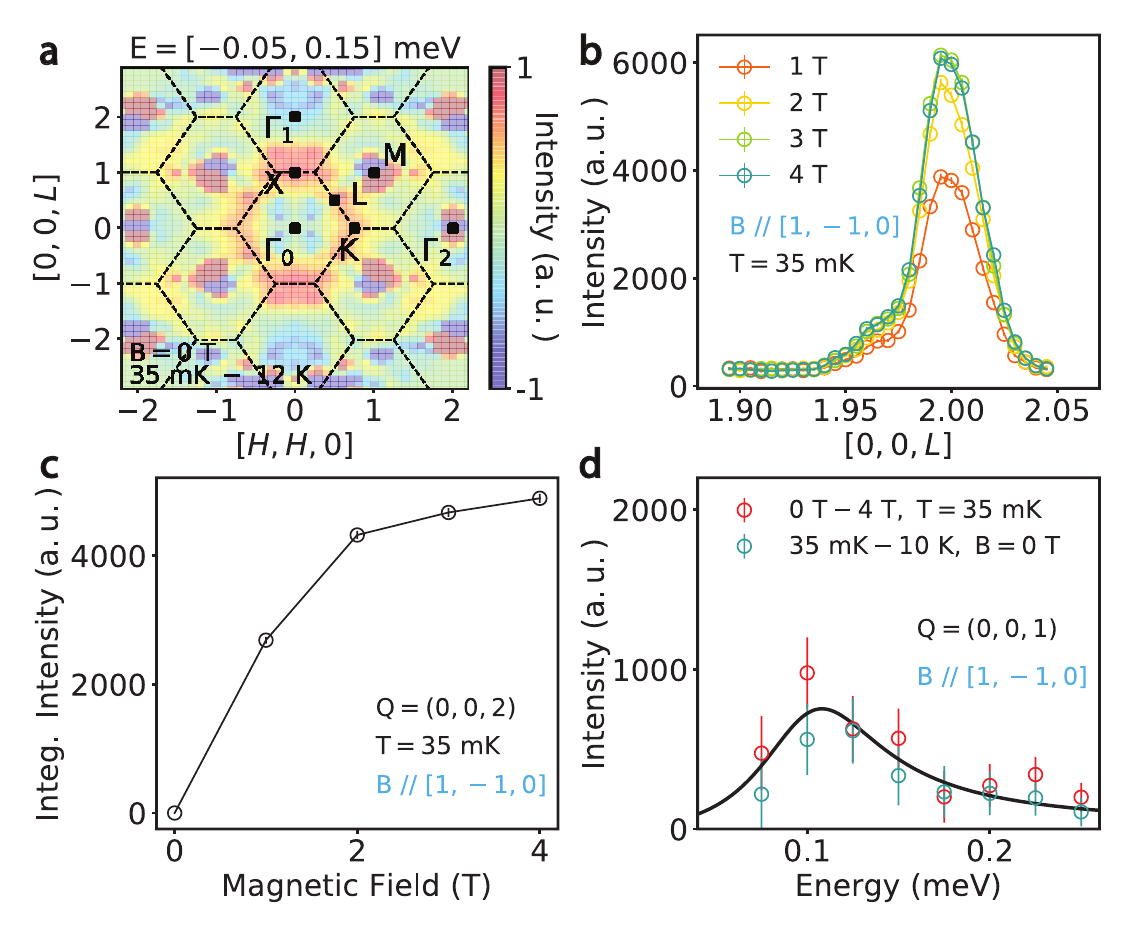}
\hl{\caption{
(a) A schematic diagram of the $[H,H,L]$ zone and the wavevector dependence of the integrated magnetic scattering from -0.05 to 0.15 meV at 35 mK, adapted from reference \cite{Gao2019}. 
(b) Wavevector cuts along the $[0,0,L]$ direction at 35 mK in 1, 2, 3, and 4 T magnetic fields. 
(c) Integrated magnetic intensity of the $(0,0,2)$ Bragg peak as a function of an applied magnetic field at 35 mK. The intensity of the Bragg peak in zero field was subtracted. 
(d) The energy dependence of the scattering obtained by subtracting 4 T data from 0 T data at 35 mK (red), and 10 K data from 35 mK data in zero field. The applied magnetic field in (b),(c) and (d) was along the $[1, -1, 0]$ direction. Data are from SPINS spectrometer.
}}
\label{Fig5}
\end{figure}

\begin{figure}[h]
\includegraphics[width=0.75\columnwidth]{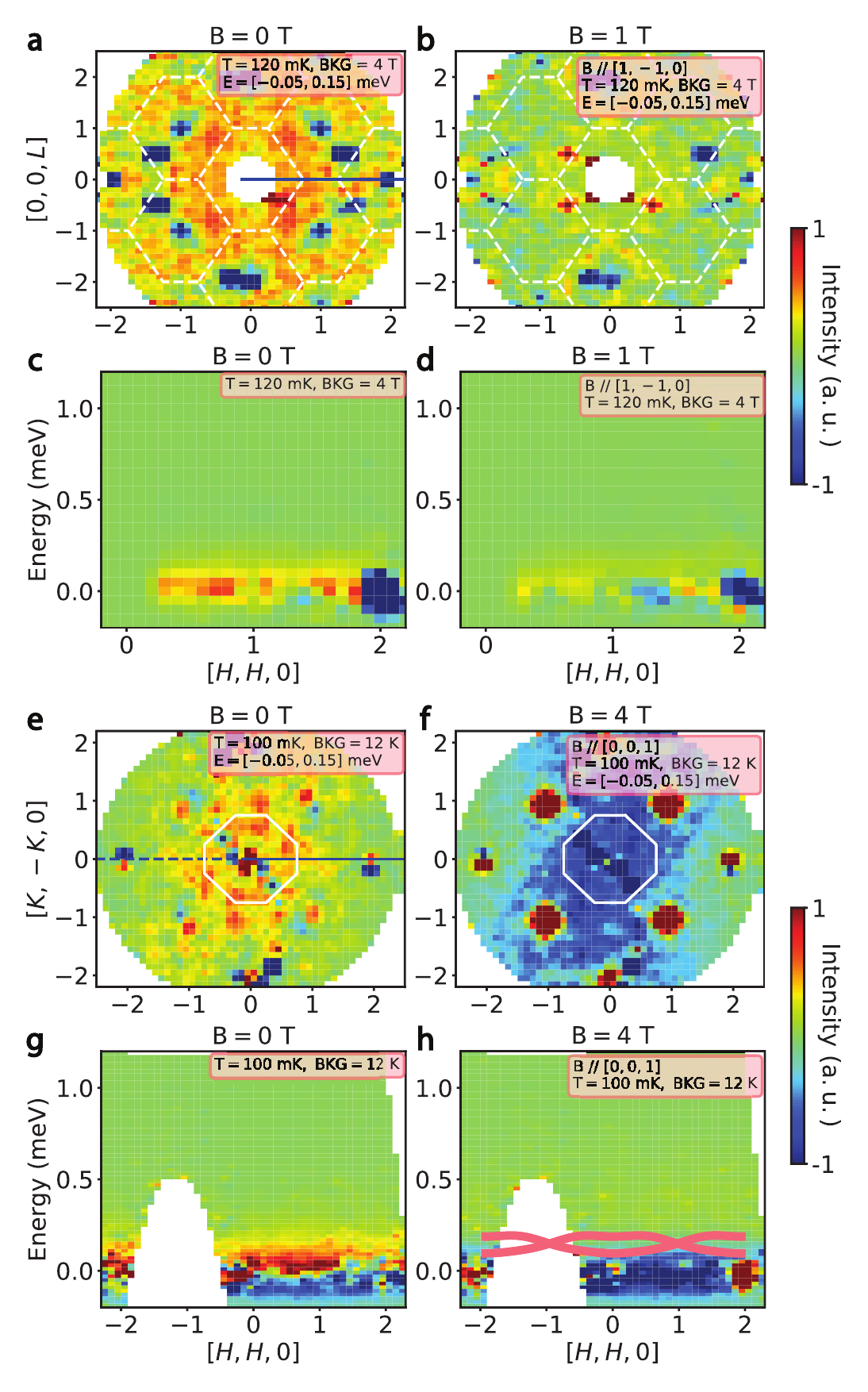}
\caption{
(a),(b) Wavevector dependence of the integrated magnetic scattering from -0.05 to 0.15 meV at 120 mK in zero and 1 T magnetic field applied along [1, -1, 0] direction. (c),(d) Dispersions for the spin excitations along [H, H, 0] direction. The spin excitation continuum in zero field almost disappears in applied magnetic fields. Data were collected from 0 to 120$^\circ$. 4 T data at 120 mK were subtracted as background. 
\bg{Data of panel a-d are from LET spectrometer.}
(e),(f) wavevector dependence of the integrated magnetic scattering from -0.05 to 0.15 meV at 100 mK in zero and 4 T magnetic field applied along [0, 0, 1] direction. (g),(h) Dispersions for the spin excitations along [H, H, 0] direction. The spin excitation continuum in zero field disappears in applied magnetic fields. Data were collected from 0 to 180$^\circ$, and expanded to 180 to 360$^\circ$ in (e), (f). 12 K data were subtracted as background. \bg{Data of panel e-h are from AMATERAS spectrometer.}
\hl{The pink curves in panel (h) highlight the position of the spin waves invisible to neutron scattering, indicating the octupolar nature of the system.}}
\label{Fig3}
\end{figure}

In this work, we report the effect of an applied magnetic field on magnetization, field-induced magnetic order, and spin excitation continuum in Ce$_2$Zr$_2$O$_7$ single crystals \cite{Gao2019}.  In addition, we carried out single crystal diffuse scattering measurements
to search for the expected octupolar excitation continuum in Ce$_2$Zr$_2$O$_7$ single crystals \cite{Sibille2020}. We find that the magnetic field directional-dependent magnetization of Ce$_2$Zr$_2$O$_7$ can be well understood by the field-induced ferromagnetic order in a QSL ground state, 
consistent with the expected magnetic structure of a field-driven Anderson-Higgs transition from the octupolar $U(1)$ QSL. 
Our inelastic neutron scattering experiments reveal that an applied magnetic field suppresses the spin excitations but does not induce ferromagnetic spin waves. 
The octupolar-character spin waves, unobservable by neutrons, nevertheless manifest their presence by clear signals in the specific heat measurements.
Finally, the angular dependence of the  diffuse scattering in our 
single crystal diffraction measurements does not exhibit strong signals at large wavevectors, in contrast with the signature seen in Ce$_2$Sn$_2$O$_7$  \cite{Sibille2020}. We reconcile this with the aforementioned octupolar QSL-like behavior by noting that Ce$_2$Zr$_2$O$_7$ has been predicted~\cite{bhardwaj2021sleuthing} to have a strong $J_x$ coupling between the dipolar $S^x$ components of Ce moment, of comparable strength to the $J_y$ coupling between the octupolar components.  This has the effect of reducing the octupolar contribution to the diffuse scattering, as observed. The present findings indicate that Ce$_2$Zr$_2$O$_7$ is a strong candidate for an octupolar QSL, 
\hl{
establishing that frustrated magnetic octupolar interactions are responsible for QSL properties in Ce-based pyrochlore magnets.}

\section{Experimental Methods}

\hl{\subsection{Sample preparation} }

Polycrystalline Ce$_2$Zr$_2$O$_7$ was synthesized using a solid-state reaction method.
Stoichiometric powders of CeO$_2$ and ZrN were mixed, ground, pelletized and sintered in a forming gas (8$\%$ H$_2$ in Ar) flow at 1,400 $^\circ$C for 20 h. 
The Ce$_2$Zr$_2$O$_7$ single crystals were grown using a floating-zone furnace at the Center for Quantum Materials Synthesis of Rutgers University. 
The elastic neutron diffraction pattern of the Ce$_2$Zr$_2$O$_7$ single crystals revealed its pure pyrochlore phase with a lattice constant of $a = 10.71$ \AA\ and good quality of the samples. The single crystals used for the experiments are well characterized by X-ray and neutron single crystal refinements, showing stoichiometric structure with about 4\% anti-site disorder between Ce and Zr (See Table S1 of Ref. \cite{Gao2019}).

\subsection{Experimental Setup and scattering geometry} 

Our neutron scattering experiments were carried out using the elastic diffuse scattering spectrometer CORELLI at the spallation neutron source, Oak Ridge National Laboratory \cite{Ye2018}, cold-neutron disk-chopper spectrometer 
AMATERAS at Japan Proton Accelerator Research Complex (J-PARC) \cite{Nakajima2011}, 
cold neutron multi-chopper spectrometer LET at ISIS Facility, STFC Rutherford-Appleton Laboratory \cite{Bewley2011},
and cold triple-axis spectrometer SPINS at NIST Center for Neutron Research. 
We define the momentum transfer $Q$ in three-dimensional reciprocal space in \AA$^{-1}$ as $\textbf{Q}=H\textbf{a}^\ast+K\textbf{b}^\ast+L\textbf{c}^\ast$, where $H$, $K$, and $L$ are Miller indices and ${\bf a}^\ast=\hat{{\bf a}}2\pi/a$, ${\bf b}^\ast=\hat{{\bf b}}2\pi/b$, ${\bf c}^\ast=\hat{{\bf c}}2\pi/c$ with  $a= b=c=10.71$ \AA\ in the $Fd\bar{3}m$ space group. 

For diffuse neutron scattering experiments, the sample was aligned in the $[H,H,0]\times[0,0,L]$ scattering plane. 
In the first experiment without applying a magnetic field, we used a $^3$He inset to regulate the temperature. 
The experiments was performed at three different temperatures, 50 K, 2 K and 240 mK, using a white incident neutron beam. 
In the second experiment with a vertical magnet, we performed scattering at 1.5 K with zero, 3 T, 4 T and 5 T magnetic field applied along the 
$[H,-H,0]$ direction.

For inelastic neutron scattering experiments on SPINS, we aligned the sample in the $[H,H,0]\times[0,0,L]$ scattering plane with vertical magnetic field along the $[H,-H,0]$ direction, and used $E_f = 3.7$ meV after the sample with an energy resolution of 0.15 meV. 
For time-of-flight neutron scattering experiments on AMATERAS, incident neutron energies of $E_i=1.7$ and 3.1 meV were used with instrumental energy resolution at elastic positions of 0.05 and 0.11 meV, respectively. 
We aligned the sample in the $[H,H,0]\times[K,-K,0]$ scattering plane with 0 T and 4 T vertical field along the $[0, 0, L]$ direction at 100 mK. 
12 K data was subtracted as the  background. 
Our assumption is that the magnetic scattering at 12 K is diffusive enough and would be wavevector/energy independent, and can thus serve as the background \cite{Gao2019}.
On LET, we used $E_i = 3.7$ meV with energy resolution of 0.13 meV. The sample was aligned in the $[H,H,0]\times[0,0, L]$ scattering plane with magnetic fields along the $[H,-H,0]$ direction at 120 mK.
Similarly, we subtracted the 4 T data as the background.

\section{Experimental and Theoretical Results}

\hl{\subsection{Field dependent magnetization} }

Figure 1(a) shows the crystal structure of Ce$_2$Zr$_2$O$_7$ where Ce and Zr tetrahedrons are marked. The magnetic Ce$^{3+}$ ions, with an effective moment of $\sim1.28\ \mu_B$  estimated from the low-temperature Curie-Weiss fit \cite{Gao2019}, form a network of corner-sharing tetrahedrons.
Because Ce$^{3+}$ ions have an effective $S=1/2$ dipole-octupole Kramer's doublet ground state, the Ce$^{3+}$ local moment has Ising-like anisotropic $g$-tensors with a parallel component (along the local $[1,1,1]$ direction of tetrahedron) $g_{||}= 2.57$ and a perpendicular component $g_\perp = 0$ \cite{Gao2019}, different from the Er$_2$Ti$_2$O$_7$ and Yb$_2$Ti$_2$O$_7$ $XY$ pyrochlores where all three components of the effective spin carry dipole moments \cite{Hallas2018}. 
Figures 1(b), 1(c), and 1(d) illustrate the expected spin configurations under applied magnetic field ${\bf B}$ along the $[1,0,0]$, $[1,1,1]$, and $[1,1,0]$ directions,
respectively. 
When ${\bf B}$ is along the $[1,0,0]$ direction, the spin configuration is 2-in-2-out, and the net magnetization $M$ along the field direction should be $M_{[1,0,0]}=1/\sqrt{3}=0.74\ \mu_B$/Ce$^{3+}$ [Fig. 1(b)].
If ${\bf B}$ is along the $[1,1,1]$ direction, the spin configuration is 3-in-1-out, and ($M_{[1,1,1]}=(1 + 3 \times 1/3)/4 =0.64\ \mu_B$/Ce$^{3+}$ [Fig. 1(c)]. Finally, when the applied field is aligned perfectly along the $[1,1,0]$ direction, two of the four spins should be perpendicular to ${\bf B}$ 
\hl{shown by the yellow arrows, and hence decouple from the magnetic field}. This would result in the expected magnetization $M_{[1,1,0]}=0.522\ \mu_B$/Ce$^{3+}$ [Fig. 1(d)]. 
\hl{The expected magnetic field-induced magnetic structure is also 3-in-1-out, although the 2-in-2-out structure cannot be ruled out based on the direction of the applied field alone [Fig. 1(d)]. }
Figure 1(e) shows the field-dependence of the magnetization along these three directions. We note that our measured values of $M_{[1,0,0]}$ and $M_{[1,1,1]}$ are below the expectation because the $B=8$~T applied field is still insufficient to saturate the moment, whereas
the measured $M_{[1,1,0]}$ is close to the saturation value of 0.522 $\mu_B$/Ce$^{3+}$
at 8~T.

\hl{\subsection{Field-induced magnetic structures} }

We first describe neutron diffraction experiments designed to determine the field-induced magnetic structure of Ce$_2$Zr$_2$O$_7$. 
For this purpose, we aligned a single crystal of Ce$_2$Zr$_2$O$_7$ in the $[H,H,0]\times[0,0,L]$ scattering plane, and applied a vertical magnetic field along the $[1,-1,0]$ direction. \hl{The alignment angles for the scattering plane are within 1$^\circ$.}
Figures 2(a) and 2(b) show 2D maps of reciprocal space in the $[H,H,L]$ scattering plane for zero and 5 T field, respectively, at 1.5 K.  
At zero field, the scattering peaks are due entirely to nuclear scattering [Fig. 2(a)]. 
The magnetic field-induced intensity gain is shown in Fig. 2(d), which reveals the magnetic Bragg peaks under a 5 T field at 1.5 K.
Since cold neutron measurements can only probe a few Bragg peaks within the scattering plane, we optioned to compare directly the field-induced integrated intensity with the magnetic structural factor calculation instead of doing a detailed refinement.
The ratio of average intensity gain of the $(2, 2, 0)$ and $(-2, -2, 0)$ vs the $(0, 0, 2)$ and $(0, 0, -2)$ magnetic Bragg peaks at 5 T is about 1.25. For the 3-in-1-out [Fig. 1(c)] and 2-in-2-out [Fig. 1(b)] magnetic structures, the expected peak intensity ratios of $(2, 2, 0)$ and $(0, 0, 2)$ are 1.4 and 0.48, respectively. 
Clearly, the 3-in-1-out spin configuration \hl{shown in Fig. 1(c)} is more consistent with the experimental results, \hl{which is also consistent with magnetization measurements of Fig. 1(e)}.  Figure 2(c) compares cuts through the Bragg peak positions illustrating the field-induced effect. 
Figure 2(e) shows the magnetic field dependence of the $(2, 2, 0)$ and $(0, 0, 2)$ intensity.
Since neutron scattering measures the square of the ordered moment, the square root of the observed magnetic Bragg intensity should agree with field dependence of the magnetization [Fig. 2(f)].

\begin{figure}[h]
\includegraphics[width=0.7\columnwidth]{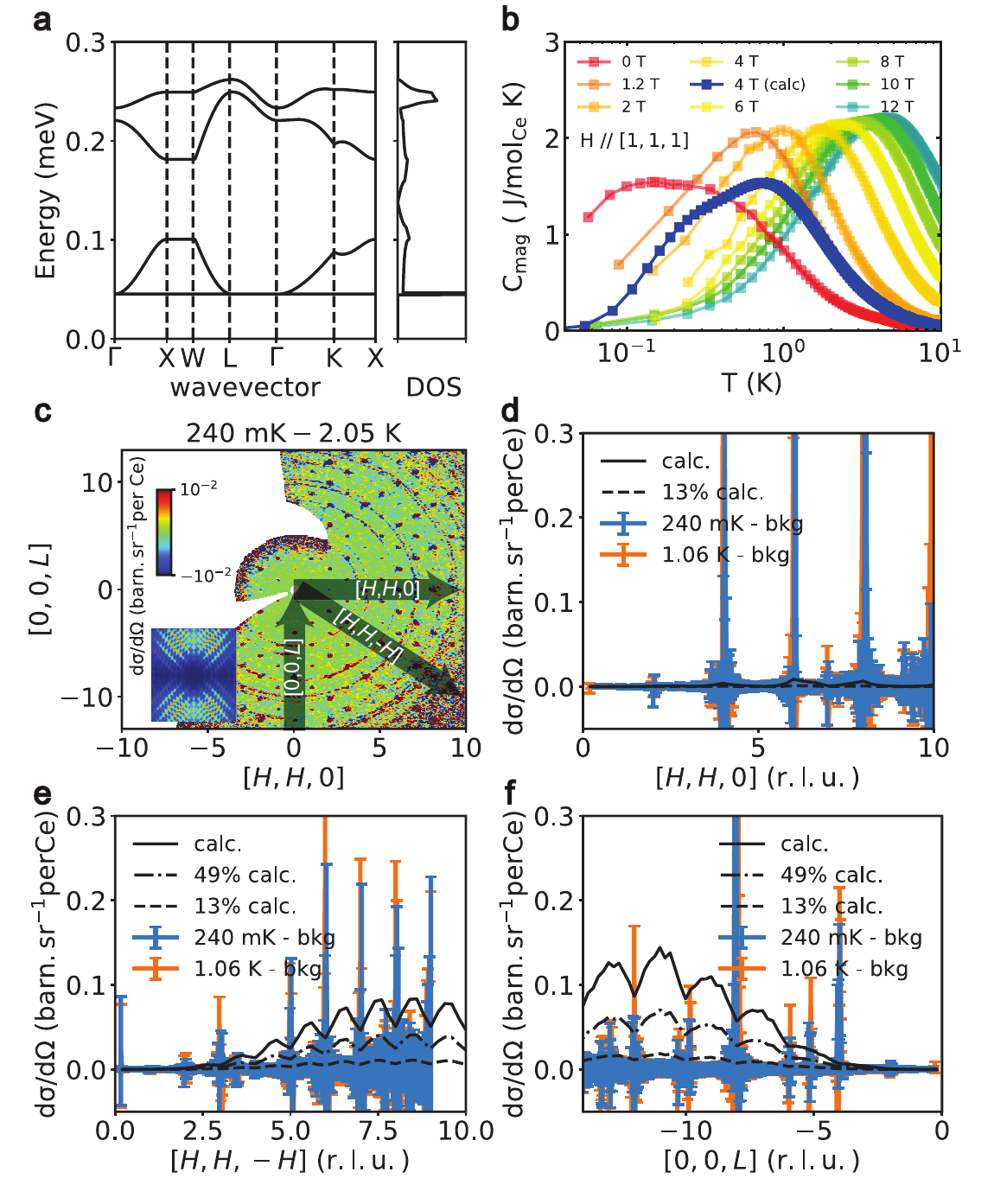}
\caption{
(a) Linear spin wave dispersion and density of state (DOS) computed with parameters in Eq.~\ref{eqn.spin.wave.parameter} and at external magnetic field $H_z=~4\text{T}$, assuming the spins are polarized in the local z-axis, forming the 3-in-1-out ground state.
(b) Specific heat measured from ref. \cite{Gao2019}, compared to the theoretical result based on linear spin wave theory.
We note that the peak positions of the theoretical curves are lower than the experimental measurements, but their overall evolution with increasing magnetic field agrees.
This indicates that the peak in specific heat does come from excitations of the spin system.
These excitations, however, are not seen in inelastic neutron scattering in Fig.~\ref{Fig3}, which can be explained by their octupole nature.
(c) Wavevector dependence of the diffuse neutron scattering at 240 mK with 2.05 K data subtracted as background. 
Arrows indicate $[H,H,0]$, $[H,H,-H]$, and $[0,0,L]$ wavevectors for cuts in (d-f). \bg{Data are from CORELLI spectrometer.}
Inset is the Monte Carlo simulations of the diffuse scattering from the octupole ice in the $[H,H,L]$ plane, adopted from reference \cite{Sibille2020} with permission.
(d-f) Wavevector cuts along the $[H,H,0]$, $[H,H,-H]$, and $[0,0,L]$ directions. The cuts on the Monte Carlo results are labeled as 'calc.'.
\hl{The vertical error bars indicate statistical errors of one standard deviation computed using
$\delta I_{(\rm 240\ mK-2.05\ K)}=\sqrt{(\delta I_{(\rm 240\ mK)})^2+(\delta I_{(\rm 2.05\ K)})^2}$, where $\delta I_{(\rm 240\ mK)}$ and
$\delta I_{(\rm 2.05\ K)}$ are statistical errors at 240 mK and 2.05 K, respectively.}
}\label{Fig4}
\end{figure}

\hl{\subsection{Effect of a magnetic field on spin excitation continuum} }

To determine how an applied magnetic field can affect the spin excitation continuum of Ce$_2$Zr$_2$O$_7$ seen in the previous work \cite{Gao2019}, 
we performed elastic and inelastic neutron scattering experiments on Ce$_2$Zr$_2$O$_7$ in the $[H,H,0]\times[0,0,L]$ scattering plane with applied field along the $[1,-1,0]$ direction using SPINS. Figure 3(a) reproduces the spin excitation continuum at 35 mK with marked high symmetry points at 35 mK, where
the 12 K data is used as background with no magnetic scattering \cite{Gao2019}.  
Figure 3(b) plots elastic scans along the $[0,0,L]$ direction at 1, 2, 3, and 4 T field at the base temperature of 35 mK. The field 
dependence of the integrated intensity is shown in Fig. 3(c), consistent with Fig. 2(e).  To determine the effect of a magnetic field
on spin excitation continuum shown in Fig. 3(a), we show in Fig. 3(d) the scattering intensity differences of
constant energy scans at 0 T and 4 T at 35 mK and $Q=(0,0,1)$ [$X$ point in Fig. 3(a)].  
The data shows a clear peak around 0.1 meV that is almost identical to the peak obtained using
temperature difference plot between 35 mK and 10 K with the same experimental setup.  
Since there is no evidence of spin excitation continuum at $\sim$10 K \cite{Gao2019}, these results
thus conclusively establish that a magnetic field of 4 T can completely suppress the continuum at 35 mK.
Therefore, the data at 4 T can be used as background for inelastic scattering.

We also carried out the inelastic neutron scattering experiments on Ce$_2$Zr$_2$O$_7$ at 120 mK with LET in the $[H,H,0]\times[0,0,L]$ scattering plane with applied field along the $[1,-1,0]$ direction. 
The incident neutron beam energy was $E_i=3.7$ meV with an energy resolution of 0.1 meV. 
Assuming that the scattering in a 4 T field is nonmagnetic in most areas in the reciprocal space except at Bragg peaks and that the applied field does not change the incoherent scattering, we can compare the spin excitation continuum in zero field [0 T $-$ 4 T, Figs. 4(a) and 4(c)] with the magnetic field-induced scattering suppression in a 1 T field [1 T $-$ 4 T, Figs. 4(b) and 4(d)].  
Figures 4(a) and 4(c) show the wavevector dependence and dispersion of spin excitations, respectively. 
At zero field, we clearly see a spin excitation continuum near the zone boundary, reproducing the previous results \cite{Gao2019}.
In the 1 T field, the spin excitation continuum is almost entirely suppressed and there is no sign of the ferromagnetic spin waves that would have been expected for a dipolar QSL. This supports the thesis that \CZO is an octupolar quantum spin liquid, proposed theoretically in this compound~\cite{bhardwaj2021sleuthing,smith2021case}.

To further test how the spin excitation continuum transforms in a magnetic field along the $[1,0,0]$ direction, we performed neutron scattering at AMATERAS by aligning the crystal in the $[H,H,0]\times[K,-K,0]$ scattering plane with the applied magnetic field along the $[0,0,1]$ direction.
Figures 4(e) and 4(f) show the wavevector dependence of spin excitations
for energies integrated from -0.05 meV to 0.15 meV at 100 mK for zero and 4 T fields, respectively. At zero field, we see a clear spin excitation continuum near the zone boundary similar to previous work in the $[H,H,L]$ scattering plane \cite{Gao2019}. When a magnetic field of 4 T is applied, the continuum disappears and eight magnetic Bragg peaks appear at $(\pm 2,0,0)$, $(0,\pm 2, 0)$, $(\pm 2,\pm 2, 0)$, and $(\pm2, \mp 2, 0)$.
A comparison of the magnetic Bragg peak intensity at these positions confirms the \hl{
2-in-2-out} field-induced structure [Fig. 1(b)].
To further determine what happens to the spin excitation continuum at 100 mK, we show in Fig.~4(g) and 4(h) the dispersions of spin excitations along the $[H,H,0]$ direction at zero and 4 T, respectively.  It is clear that a 4 T field suppresses the spin excitation continuum but does not induce spin waves below 1.2 meV, corroborating the octupolar nature of the excitations.

\hl{\subsection{Effect of magnetic field and absence of magnons in neutron scattering} }
The octupolar character of the spins can also be deduced by contrasting 
the spin waves  [Fig.~5(a)]  and specific heat [Fig.~5(b)] with neutron scattering 
measurements [Figs.~3(f)] \hl{\cite{Gao2019,smith2021case}.}
At external field $B\ge 4$~T in the $[1,1,1]$ direction,
we see a significant specific heat signal with a peak evolving from  $\sim1.7$ K at $B=4$ T to $\sim 4.9$ K at $B=14$ T.
\hl{
Although we have not performed inelastic neutron scattering experiments for field along the $[1,1,1]$ direction, it is clear
that such measurements would not yield any spin wave signal in the expected energy range based on our measurements
for fields along the $[1,-1,0]$ and $[1,0,0,]$ directions discussed in Figs. 2-4.}

From the magnetization curve in Fig.~1(e), we know that at external field  $B\ge4$ T,
the spins are \hl{almost} saturated in the local $S^z$ direction, forming   ``3-in-1-out'' or ``3-out-1-in'' configurations on the two types of tetrahedra [Fig.~(1c)]. 
One expects the excitations in this case to be well-defined spin waves.
Indeed, the magnon dispersions [Fig.~5(a) left panel], density of states [Fig.~5(a) right panel], and the resulting 
specific heat [dark blue squares in Fig.~5(b)]
can be analytically calculated.
Using the parameters 
\begin{equation}
	\label{eqn.spin.wave.parameter}
	\begin{split}
		&J_x = J_y = 0.068~\text{meV},\;
		J_z = 0.013~\text{meV}, 
		\hy{J_{xz} = 0}\\
		&g_z = 2.3,\;
		g_x=g_y = 0 
	\end{split}
\end{equation}

\bg{in the Hamiltonian
\begin{equation}
		{H}_{{\mathrm{nn}}}=\mathop{\sum}\limits_{\langle ij\rangle }{J}_{y}{s}_{i}^{y}{s}_{j}^{y}+[{J}_{x}{s}_{i}^{x}{s}_{j}^{x}+{J}_{z}{s}_{i}^{z}{s}_{j}^{z}+{J}_{xz}({s}_{i}^{x}{s}_{j}^{z}+{s}_{i}^{z}{s}_{j}^{x})]
\end{equation}}

\begin{figure}[th]
	\includegraphics[width=0.5\columnwidth]{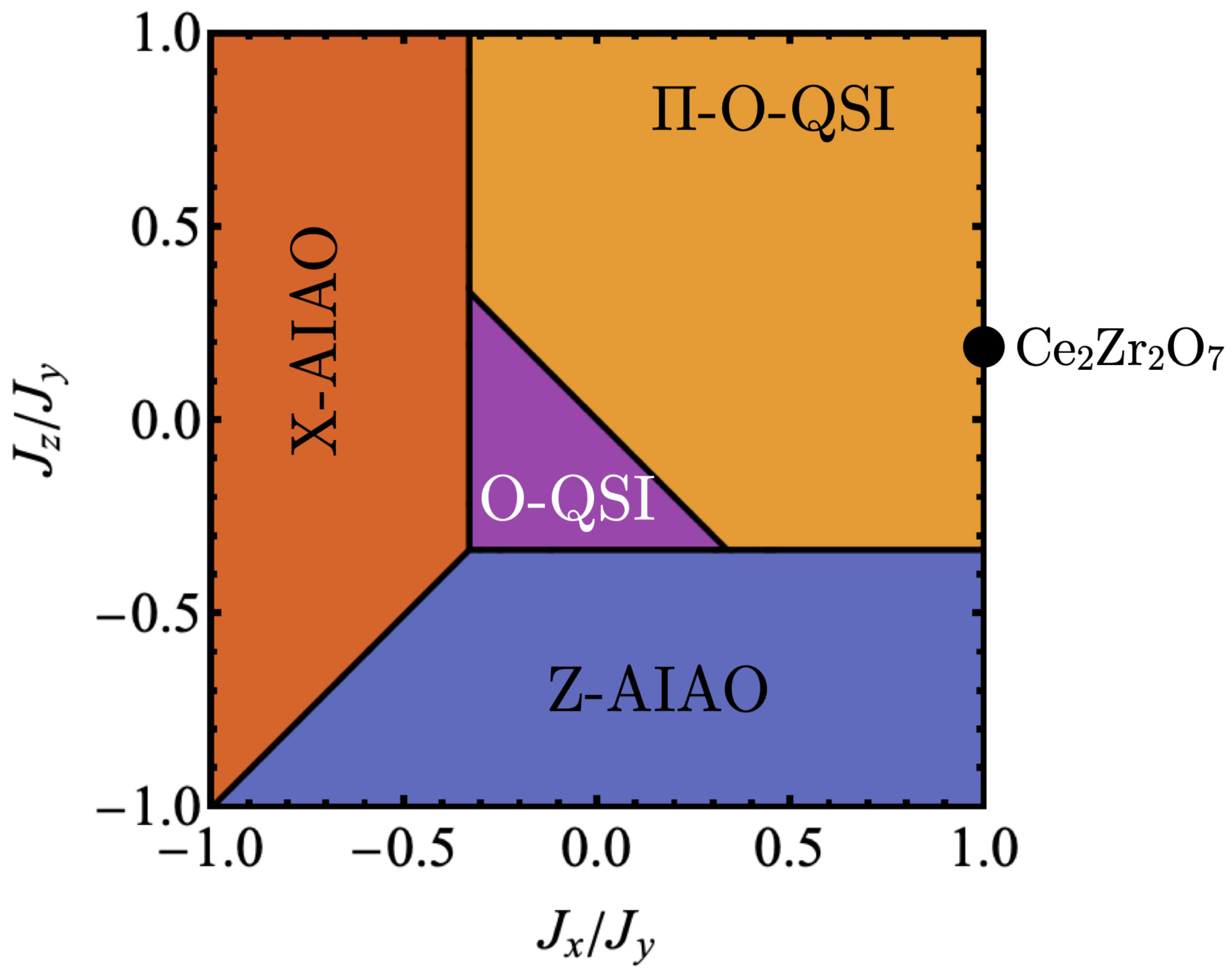}
	\caption{\hy{The parameters in Eq.~(\ref{eqn.spin.wave.parameter}) places  the model of   \CZO\ in the $\pi-$flux   quantum spin ice phase. The phase diagram is from Ref.~\cite{Patri2020}.}}
\label{Figure_phase_diagram}
\end{figure}

estimated  from two  independent and mutually consistent studies by Bhardwaj \textit{et al.}~\cite{bhardwaj2021sleuthing} and  Smith \textit{et al.}~\cite{smith2021case}.
At zero external field, these parameters place the model in the $\pi$-flux octupolar spin liquid phase \cite{Patri2020}  [Fig.~\ref{Figure_phase_diagram}].
At finite magnetic field, we found that 
\hl{
the spin waves,
whose band dispersion is distributed around $0.05- 0.25~$ meV [Fig.~5(a)], 
 can qualitatively explain the specific heat peaks measured at $B\ge4$ T.}

\begin{figure}[th]
	\includegraphics[width=\columnwidth]{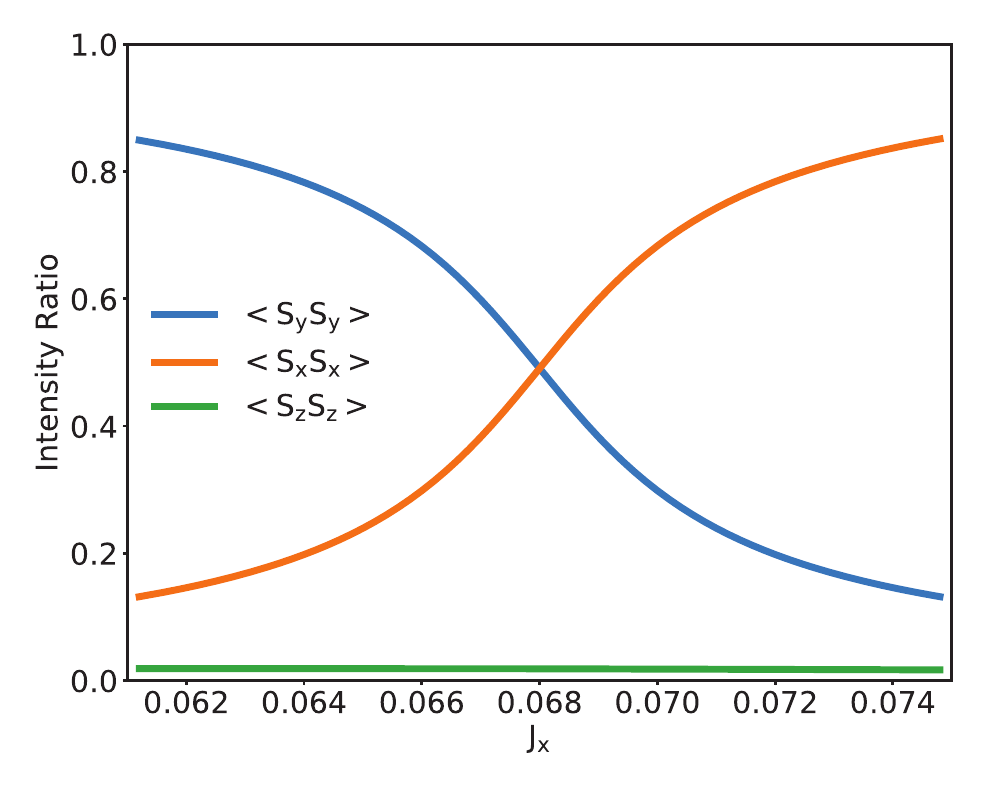}
	\caption{\hl{Correlation intensity in the octupole-octupole channel $(\langle S^y S^y \rangle)$ as well as the dipole-dipole channels  $(\langle S^x S^x\rangle,\langle S^z S^z\rangle )$ when varying $J_x$ and keeping $J_y,\ J_z$ in Eq.~\eqref{eqn.spin.wave.parameter} fixed.
	These parameters are estimated to represent the model of \CZO.
	The intensity is computed using Self-Consistent Gaussian Approximation.
	The $\langle S^y S^y \rangle$ channel has around 50\% of its highest intensity (the case of \CSO) when $J_x = J_ y \gg J_z$.
	As $J_x$ increases slightly by 10\%, the $\langle S^y S^y \rangle$ channel intensity drops fast to only 13\%.
	Since the neutrons at large momentum only couple to  the $\langle S^y S^y \rangle$, 
	it is very plausible that large $J_x$ reduces its signal to be beyond the limit of detection}.
	}
\label{Fig6}
\end{figure}

Using the same parameters, we can also compute the spin-wave when
the system is nearly saturated in the $[1,0,0]$ external field of $4$ T [Fig. 4(h)], 
and form a ferromagnetic 2-in-2-out ordered ground state.
Again, the magnon dispersions are at a similar energy scale of $0.05- 0.2~$ meV [Fig. 4(h)].
However, they are not observed by inelastic neutron scattering technique, although  
this energy range well accessible with the experiment [Fig. 4(h)]. \bg{One may argue that the intensity of spin waves
may be too weak to be detected. In our inelastic neutron experiments at both LET and AMATERAS, we counted for one whole day for each temperature (field), which is much longer than the normal scan time for spin wave measurement ($4 \sim 6$ hours). 
}

This seeming contradiction between the two different experiments can be resolved by noting the octupolar nature of the Ce magnetic moments. 
In the high external field regime, the magnon excitations are created by $S^\pm = S^x\pm iS^y$, and hence are expected to appear in the spin-spin correlation channels of $\langle S^x S^x \rangle$, $\langle S^y S^y \rangle$ in the neutron scattering experiments.
However, due to the octupolar nature of the spins, the $\langle S^y S^y \rangle$ channel does \textit{not} couple to the neutrons at the small momentum range measured, 
while the $\langle S^x S^x \rangle$ channel is also invisible due to the vanishing or very small $g_x\approx 0$.

\subsection{Octupolar v.s. dipolar quantum spin ice phase}
Finally, we comment on the distinction between the dipolar and octupolar $\pi-$flux quantum spin ice phases.
	The dipolar phase means the dipole component $S^x$ or $S^z$ plays the 2-in-2-out spin ice role while other terms give rise to quantum dynamics to the ice-states. 
	The octupolar phase means the component $S^y$ plays the  spin ice role. 
	When the ice-rule enforcing terms are order of magnitude larger than the quantum dynamics terms, this is a legitimate viewpoint.
	However, the situation is more subtle in our case, when $J_x$ is comparable to $J_y$. 
	Theoretically, it is still an open question how phase transition(s) happens near $J_x = J_y$. At large $J_x$ or $J_y$, it is fairly convincing that the system is described by the dipolar and octupolar electrodynamics, respectively, which are distinct by symmetry enrichment. But it is not clear whether these are the only two phases in the parameter space, or if there are other spin liquid phases in-between. This also means that the phase transition procedure between the two phases is not well-understood yet.

It is beyond the scope of this experimental work to address this question, and has no definitive theoretical answer. 
	It is often assumed that the system is dipolar when $J_x> J_y$ and octupolar when $J_y> J_x$, 
	but the transition between the two phases has not been investigated \cite{Benton2020,Patri2020,smith2021case}.
	Mean-field study of the same Hamiltonian instead suggests that the entire $\pi-$flux phase with positive $J_x, J_y, J_z$ is one phase without phase transition  when one coefficient becomes larger than the another \cite{LeePRB2012}. 
	The exact diagonalization results of Ref. ~\cite{Patri2020} do not observe a phase transition when $J_x$, $J_z$ are 
	comparable to $J_y$. Unfortunately, this is close to the edge of the phase diagram and the differences between different theoretical investigations seem to suggest a more complicated situation.

If the  dipolar and octupolar $\pi-$flux quantum spin ice phases are indeed two distinct phases with a quantum phase transition separating them, 
then the current parameters place \CZO\ very close to the phase boundary and further investigation is needed 
to conclusively determine its ground state. In the more complex scenario, \CZO\ is still an interesting material in the $\pi-$flux quantum spin liquid phase with a significant portion of octuplar quantum fluctuation, but the precise nature of its phase awaits more investigation.

\hl{\subsection{Search for octupolar scattering}}

\hy{ This also addresses the difference between the neutron diffuse scattering experiments on  Ce$_2$Zr$_2$O$_7$ and its sister compound Ce$_2$Sn$_2$O$_7$, which is suggested to be a octupolar quantum spin liquid  \cite{Sibille2020}.}
Recently, neutron diffuse scattering experiments on the powder samples of a sister compound Ce$_2$Sn$_2$O$_7$ 
revealed a broad peak at large wavevectors (5-10 \AA$^{-1}$) \cite{Sibille2020}.  The position and intensity of the peak are consistent with the powder averaged Monte Carlo simulations of the classical octupole ice \hl{in absolute intensity units,} instead of 
the spin ice of magnetic dipoles [Fig. 5(c) inset] \cite{Sibille2020}. 
Motivated by their work, we have performed diffuse neutron scattering on Ce$_2$Zr$_2$O$_7$, with the resulting signal shown in Figure 5(c) at 240~mK, with the background at 2.05~K subtracted, in absolute units (see appendix). In contrast to Ce$_2$Sn$_2$O$_7$, we did not observe a clear signature of octupole scattering at the large wavevector region, demonstrated in Figures 5(d), 5(e), and 5(f) which show cuts along the high-symmetry directions $[H, H, 0]$, $[H, H, -H]$, and $[0, 0, L]$, respectively.
\hl{The large error bars at Bragg peak positions are due to large intensity of nuclear Bragg peaks.}
The signals \hl{at other wavevectors} are much weaker at large momentum compared with even 25\% of the Monte Carlo simulations within the statistical errors of the measurements, in contrast to the case in \CSO\  ~\cite{Sibille2020},
or the neutron scattering prediction  of uncorrelated cerium ions \cite{Lovesey2020}. 

An explanation consistent with other experimental results is \hl{
based on}  a significant $J_x \sim J_y \gg J_z$ in \CZO~ \cite{bhardwaj2021sleuthing,smith2021case} 
while in \CSO\ only $J_y$ is dominant~\cite{Sibille2020}.
As a result, in \CSO, the spins could be thought (in a \hy{
quantum, perturbative  picture}) as being mostly aligned in the octupole $S^y$ direction,  
hence producing the strong octupole-octupole correlations \hl{$(\langle S^y S^y \rangle)$} measured by diffuse neutron scattering at large $|Q|$.
In \CZO, however,  $J_x$  and $J_y$ are estimated to be on the same scale, 
\hl{
which reduces the neutron signal strongly, as the spins are also inclined to point in the $S^x$ direction.
In particular, when $J_x$ becomes greater than $J_y$, the 
octupole-octupole correlation intensity should drop fast. 

The quantitative  calculation of the signal intensity can be made using Self-Consistent 
Gaussian Approximation (SCGA) \cite{Isakov2004PhysRevLett}.
The details of this method is given in the Appendix, and the results are shown in Fig.~\ref{Fig6}.
The parameters given in Eq.~\eqref{eqn.spin.wave.parameter}
has $J_x = J_y \gg J_z$.
The expected diffuse neutron scattering strength is then
only $49\%$ of that of \CSO. 
Furthermore, a small increase in $J_x$,
well within the reasonable range of estimation in Ref.~\cite{bhardwaj2021sleuthing}, 
will further reduce the expected neutron scattering strength.
For example, a 10\% increase in $J_x$ to
 $J_x = 0.075$
 reduces the $\langle S^y S^y \rangle $ to only $13\%$ of \CSO, 
making it hardly detectable in the experiment.
Hence, the absence of neutron scattering at large momenta is not surprising, and actually highly possible with previous theoretical works suggesting a large magnitude of $J_x$ (comparable to $J_y$) in \CZO\ \cite{bhardwaj2021sleuthing,smith2021case}. }

\hl{\section{Conclusions}}

In summary, we use the elastic and inelastic neutron scattering measurements to identify the Ce$_2$Zr$_2$O$_7$ system as an octupolar $U(1)$ QSL, in which the application of a magnetic field induces an Anderson-Higgs transition by condensing the spinons into the static ferromagnetic order, however with the associated spin waves invisible to neutrons due to the octupolar nature of Ce spins. 
These octupole spin waves however have a clear signature, and can be quantitatively accounted for, in our specific heat measurements.
Furthermore, we demonstrate that the diffuse neutron scattering at large momenta expected of \hl{ an octupolar order} is nearly absent. We offer an explanation of this puzzling phenomenon, which lies in the considerable magnitude of the quantum 
\hl{$J_x  S_i^x S_j^x$} terms in the effective spin ice model, consistent with the previous theoretical studies.    
\hy{This finding indicates that Ce$_2$Zr$_2$O$_7$ is a strong candidate of octupolar spin liquid. 

}

\section*{Acknowledgements}

We are grateful to Romain Sibille and Petit Sylvain for providing us with raw data of powder
results on Ce$_2$Sn$_2$O$_7$ and the Monte Carlo simulations for the octupole ice shown
in Fig. 4(c). 
We thank Arthur Ramirez, Owen Benton, and Collin Broholm for helpful discussions. 
The neutron scattering work at Rice is supported by US DOE BES DE-SC0012311 (P.D.). The theoretical work at Rice was supported by the National Science Foundation Division of Materials Research Award DMR-1917511 (H.Y. and A.H.N.). 
The single-crystal-growth work at Rice is supported by the Robert A. Welch Foundation under grant no. C-1839 (P.D.). E.M. and C.-L.H. acknowledge the support from US DOE BES DE-SC0019503. Crystal growth by B.G. at Rutgers was supported by the 
visitor program at the center for Quantum Materials Synthesis (cQMS), funded by the Gordon and Betty Moore Foundation’s EPiQS initiative through grant GBMF10104, and by Rutgers University. Polycrystalline preparation by X.X. was supported by the DOE under Grant No. DOE: DE-FG02-07ER46382.

\appendix

\section{Neutron Scattering Intensity in Experiment}

To directly compare the reported octupolar scattering as the diffusive signal at high wave vectors in Fig. 4(c) of main text with our diffuse neutron scattering experiments on CORELLI, we have to convert the diffuse scattering in our data to absolute units. The definition of the differential cross section is \cite{lovesey1984}:
\begin{equation}
\frac{d \sigma}{d \Omega} = \frac{N}{\Phi d \Omega} = \frac{\sum_i N_i}{\sum_i (\phi_i  d \Omega_i)}
\end{equation}
where $N$ is the number of scattered neutrons per unit time in an infinitesimal volume ($d Q$) of reciprocal space, around a momentum transfer ($Q$), $\Phi$ is the incident flux and $d \Omega$ is the solid angle of the detector. The $\sum_i$ in the last step is for experiments using multiple detectors or polychromatic incident beams. 

To obtain absolute scattering intensity, the standard procedure is to measure the incoherent scattering of a vanadium standard using the same experimental setup. Since the vanadium incoherent scattering is isotropic, the differential scattering cross section is written as:
\begin{equation}
\frac{d \sigma}{d \Omega} = \frac{\sigma_I}{4 \pi}
\end{equation}
where $\sigma_I$ is the total incoherent scattering cross section. Then we will have 
\begin{equation}
\sum_i (\Phi_i d \Omega_i) = \sum_i V_i / \frac{\sigma_I}{4 \pi}
\end{equation}
where $V_i$ are the neutron counts from vanadium. Thus,
\begin{equation}
\frac{d \sigma}{d \Omega} = \frac{\sigma_i}{4 \pi} \frac{\sum_i N_i}{\sum_i V_i}
\end{equation}
The $N_i$ and $V_i$ are corrected for the absorption of the sample and vanadium, respectively. The standard procedure mentioned above was used to treat the data on CORELLI with the MANTID (Manipulation and Analysis Toolkit for Instrument Data) program \cite{arnold2014}. 

From ref. \cite{michelsclark2016}, the integrated intensity of a Bragg peak is given by:
\begin{equation}
I_c = V N(\lambda) \frac{\lambda^4 |F(\tau)|^2}{2 v_c^2 sin^2(\theta/2)}
\end{equation}
where $V = N v_c$ is the sample volume, $N$ is the number of coherent scatters, $v_c$ is the unit-cell volume, $F(\tau)$ is the unit-cell structure factor, and $\theta$ is the conventional polar angle of a spherical coordinate system, not the crystallographic angle $2 \theta$.

Similarly, the incoherent scattering intensity is given by:
\begin{equation}
I_i = N(\lambda) \frac{N_i \Delta {\bf Q} \sigma_i}{4 (2 \pi)^4} \frac{\lambda^4}{2 sin^2(\theta / 2)}
\end{equation}
where $\Delta {\bf Q}$ is the integration volume element and can be a user-defined constant in the formula.
Since the integration is defined to be over the same small volume for the Bragg peak (equation (5)) and the incoherent scattering (equation(6)), the flux term and Lorentz factor are identical, yielding
\begin{equation}
|F(\tau)|^2 = c I_c / I_i
\end{equation}
where $c$ is a wave vecter and detector independent constant. Note that although the standard normalization procedure described above should have errors less than 10$\%$, it is often less accurate in practice. 

\hl{\section{Self-Consistent Gaussian Approximation }}

\hl{The Self-Consistent Gaussian Approximation 
is an analytical method that treats
the spin in the Large-N limit.
Our calculation follows closely 
Ref.~\cite{Isakov2004PhysRevLett}.
This method treats $S^{x,y,z}$ as independent, freely fluctuating degrees of freedom,
except for a Lagrangian multiplier term that enforces the averaged spin norm condition.

The Hamiltonian in momentum space is written as
\begin{equation}
	\mathcal{E}_\text{Large-N} = 
	\frac{1}{2}\mathbf{S}\mathcal{H}_\text{Large-N} \mathbf{S}^T ,
\end{equation}
where $\mathbf{S} = (S_1^x, S_2^x, S_3^x, S_4^x,\dots,  S_3^z, S_4^z).$
The interaction matrix ${H}_\text{Large-N}$
is the Fourier transformed interaction matrix that includes the nearest neighbor interactions of $J_x,\ J_y,\ J_z$.
In our case, 
it decouples into three block diagonal matrices for $S^x$, $S^y$, $S^z$ interactions separately,
\begin{equation}
{H}_\text{Large-N} = \begin{pmatrix}
	\mathcal{H}_x & 0 & 0 \\
	0 & \mathcal{H}_y & 0\\
	0 & 0 &  \mathcal{H}_z
\end{pmatrix}.
\end{equation}
Each block is of form
\begin{align}
&\mathcal{H}_\alpha = 2J_\alpha\times \\
&\begin{pmatrix}
	0 & \cos(q_y + q_z) & \cos(q_x + q_z)&\cos(q_x+ q_y) \\
	\cos(q_y + q_z) & 0 &  \cos(q_x - q_y) & \cos(q_x - q_z) \\
	\cos(q_x + q_z) & \cos(q_x - q_y) & 0 & \cos(q_y - q_z) \\
	\cos(q_x+ q_y) & \cos(q_x - q_z) & \cos(q_y- q_z) & 0
\end{pmatrix},
\end{align}
where $\alpha = x,\ y,\ z$.

We then introduce a Lagrangian multiplier
with coefficient $\mu$
to the partition function
to get
\begin{equation} 
	\mathcal{Z} =   \exp\left( -\frac{1}{2} {\int_\text{BZ}  \text{d}\bfq\  \text{d}\mathbf{S} \  \mathbf{S}  \left[\beta  \mathcal{H}_\text{Large-N}+ \mu
		\mathcal{I}\right]\mathbf{S} }  \right)   
\end{equation}
in order to impose an additional constraint
of averaged spin-norm being $S^2$, or
\begin{equation}
	\langle  \mathbf{S}_1^2  +  \mathbf{S}_2^2 +   \mathbf{S}_3^2 +   \mathbf{S}_4^2 \rangle  = 1 .
\end{equation}
For a given temperature $k_B T= 1/\beta$, the value of $\mu$ is fixed by this constraint via relation 
\begin{equation}
	\int_\text{BZ}   \text{d} \bfq \sum_{i=1}^{12}\frac{1}{\lambda_i(\bfq)+\mu} = \langle \mathbf{S}_1^2 + \mathbf{S}_2^2 +\mathbf{S}_3^2 +\mathbf{S}_4^2   \rangle = 1   ,
\end{equation}
where $\lambda_i(\bfq),\ i=1,2,\dots, 12$
are the twelve eigenvalues of $\beta\mathcal{H}_\text{Large-N}$.
With $\mu$ fixed,
the partition function 
is completely determined for a free theory
of $\mathbf{S}$,
and all correlation functions can be computed from $\left[\beta  \mathcal{H}_\text{Large-N}+ \mu
\mathcal{I}\right]^{-1}$.

In particular,  for a given $\alpha = x,\ y,\ z$,
\begin{equation}
\langle S^{\alpha }S^{\alpha }\rangle \sim   \int_\text{BZ}   \text{d} \bfq \sum_{i=1}^{4}\frac{1}{\lambda^\alpha_i(\bfq)+\mu},
\end{equation}
where $\lambda^\alpha_i,\ (i=1,2,3,4)$ are the eigenvalues of 
the  block $\mathcal{H}_\alpha$.
This allows us to compute the strength of different channels $\langle S^{\alpha }S^{\alpha }\rangle$.}

\appendix 

\end{document}